# Evidence that the upper critical field of Nb$_3$Sn is independent of whether it is cubic or tetragonal

Jian Zhou, Younjung Jo*, Zu Hawn Sung, Haidong Zhou, Peter J. Lee, David C. Larbalestier

*National High Magnetic Field Laboratory, Florida State University, Tallahassee, Florida 32310, USA*
*\*Now at Department of Physics, Kyungpook National University, Daegu 702-701 South Korea*

**Abstract:**

Although 2011 marks the 50th anniversary of Nb$_3$Sn as the first high field superconductor, real understanding of its upper critical field behavior $\mu_0H_{c2}$ is incomplete. Here we show surprising $\mu_0H_{c2}$ data on highly homogeneous bulk samples examined both by small-current, transport and by volumetric-averaging specific heat and the reversible magnetization techniques, which exhibit identical upper critical field $\mu_0H_{c2}(0.3\ K) \sim 29\pm 0.2$ T with or without undergoing the cubic to tetragonal transition, a result in strong contrast to widely used multiple-source data compilations that show a strong depression of $\mu_0H_{c2}(0K)$ from 29 T to 21.4 T in the tetragonal state.

Nb$_3$Sn is the most widely used superconductor for generating fields above ~10 T because it is easily and economically fabricated in round-wire, multifilament forms that lend themselves both to laboratory magnets and to the cabled, high-current conductors needed for accelerator and fusion uses. More than 600 tons of Nb$_3$Sn will be used in the International Tokomak Experimental Reactor (ITER). This wide and long-standing use [1] makes it all the more surprising that there is no agreed data set that shows the variation of the upper critical field $\mu_0H_{c2}$ across the variable composition of the A15 phase of Nb$_3$Sn. High Sn compositions are important for the very high critical current densities $J_c$ now achieved in commercial strands.[2] Optimizing $J_c(H)$ would be much easier with a quantitative understanding of how $\mu_0H_{c2}$ varies with Sn content since all practical Nb$_3$Sn wire forms contain the full range of A15 phase compositions (generally thought to range from ~18-25 at% Sn [3]). The integrated effect of shells of varying composition and thus $T_c$ and $H_{c2}$ has been addressed both experimentally [4] and by modeling [5]. These studies show that flattening the Sn gradient raises the effective $\mu_0 H_{c2}$ and $J_c$. The modeling scheme of Cooley *et al.* [5] is particularly valuable, but it lacks key information, namely the composition dependence of $\mu_0 H_{c2}$. To remedy this lack, we have fabricated homogeneous binary $Nb_3Sn$ bulk samples to provide the compositional variation of $\mu_0 H_{c2}$ with one sample set made in one, consistent fashion. Here we show whether the sample transforms to the tetragonal state or not is irrelevant to $\mu_0 H_{c2}(0)$, which can equal 29.1 ± 0.2 T in both cases, values as high as any optimally Ti- or Ta-doped wire [4] [6].

Rather homogeneous bulk $Nb_3Sn$ was made by Devantay *et al.* [7] by heating samples into the melt phase and by Goldacker *et al.* [8] in a Hot Isostatic Press (HIP) at 1100 °C. To further reduce inhomogeneity, we used a HIP capable of reaching up to 2200 °C. Here we report on bulk $Nb_3Sn$ with nominally stoichiometric 25at% Sn and Sn-rich 27at% Sn so as to ensure the most Sn-rich composition of the A15 phase. About 45 g samples were synthesized by combining Nb (-325 mesh, 99.8%, Alfa Aesar) and Sn (-325 mesh, 99.8%, Alfa Aesar) powders in a high energy ball mill. Mixing and powder packing was performed in a dedicated glove box filled with Ar gas to minimize oxidation. After 60 minutes of ball milling, the mixed powders were pressed in a Cold Isostatic Press (CIP) to form a hard pellet, then wrapped in Ta foil and put into a steel tube with one closed end. This HIP tube was then evacuated and the open end sealed by welding. The sealed can was pressurized at 2 kbar during both a pre-anneal at 650 °C for 16 hr and during the main A15 phase reaction at 1200 °C for 72 or 160 hr. The central reacted A15 part of each of the cans was cut into 2 pieces using a precision diamond saw, one piece being then re-sealed in an evacuated Ta-lined Nb tube for a 2[nd] HIP homogenization and reaction-continuation anneal at 1400 °C, 1600 °C or 1800 °C for 24 hr. One piece of the 27 at% Sn sample annealed at 1800°C was further annealed in a Ta-lined Nb tube for 30 days at 1200 °C. In this report we describe samples by their nominal or overall atomic %Sn content, followed by the final heat treatment and time (if not specified, 24 hr). For example, 25Sn_1800 means the sample finally annealed at 1800 °C for 24 hr after pre-annealing at 650 °C for 16 hr and reaction at 1200 °C for 72 hr. However, we also report the measured compositions of the A15 grains in each sample when considering the final properties.

X-ray diffraction (XRD) was used to measure lattice parameters and search for the tetragonal transformation in an instrument equipped with a helium cryostat. Specific heat measurements of the $T_c$ distribution were performed in a 16 T Physical Property Measurement Systems (PPMS). $H_{c2}(T)$ was mostly determined by small-current resistivity measurements in fields up to 32 T down to 0.3 K but was benchmarked in several cases by measurements of $\mu_0 H_{c2}$ derived from the reversible magnetization in a 14 T vibrating sample magnetometer. The samples were so well annealed that they showed very little magnetization hysteresis and clear $H_{c2}$ transitions. A15 grain compositions were determined using pure-element-standardized energy dispersive X-ray spectroscopy (EDS) in a field emission Scanning Electron Microscope (SEM).

Fig. 1 shows the resistive and specific heat superconducting transitions and the calculated $T_c$ distributions [9] [10] for both compositions in their differently heat treated conditions. Table 1 summarizes the $T_c$, lattice parameters $a_0$, the EDS-measured Sn at%, the Residual Resistance Ratio (RRR) [=$\rho(300 K)/ \rho(20 K)$], $\rho(20 K)$ and the resistively measured $\mu_o H_{c2}$ at 0.3 K. Raising the annealing temperature to 1800 ºC reduces the A15 Sn content by about 1at.% in each case, but $T_c$ drops only slightly on raising the annealing temperature from 1400 °C to 1800 °C. All the data indicate that Sn is rejected from the A15 lattice above 1400 °C, consistent with the Nb$_3$Sn phase boundary bending to Sn-poor compositions as in published Nb-Sn phase diagrams [3, 11]. High sample homogeneity is attested by narrow resistivity transitions (except for the 27Sn_1400 sample) and especially by the narrow specific heat $T_c$ distributions of Fig 1. Moreover, high homogeneity is also evidenced by the fact that the volumetric averaging measurement of the reversible magnetization yields sharp magnetization slope transitions at $H_{c2}(T)$ which overlap very closely with the potentially percolative, "best bit" resistivity measurements (Fig. 2). We thus have good confidence that the $\mu_0 H_{c2}(T)$ values from our high-field resistivity measurements represent the entire sample, except in the one sample 27Sn_1400 shown in Fig. 1.

The cubic to tetragonal structural transformation of the A15 phase has been reported to occur at temperatures varying from 31 K to 45 K [12]. For example, the tetragonal transformation takes place below 45 K in the study of Devantay *et al.* [7] and at 35 K in the samples of Goldacker *et al.* [8], when Sn is >24.5% [8]. Fig. 3 shows the {400} XRD peaks obtained at room temperature and at 10 K for both compositions in various conditions. The split of the 2θ peak around 71.4° at room temperature into two peaks at 10 K is clear evidence of the cubic to tetragonal transformation. The contraction along the c-axis distinguishes the (004) plane reflection from the {400} family and moves it to a higher angle. While 25Sn_1400, 25Sn_1600, 27Sn_1400, and 27Sn_1200_30days display the tetragonal structure in Fig. 3, 25Sn_1800 and 27Sn_1800 remain cubic down to 10 K. The EDS-measured Sn contents of our tetragonal samples are 24.6% ("25Sn_1400"), 24.5% ("25Sn_1600"), 24.6% ("27Sn_1400"), and 24.7% ("27Sn_1200"), respectively, while those of the cubic samples are 23.3% ("25Sn_1800") and 23.7% ("27Sn_1800"). This result does conform that the cubic to tetragonal transformation occurs only for Sn higher than 24.5%Sn [7] [8] [13].

Fig. 4 (a) shows $\mu_0H_{c2}(T)$ of 25Sn_1400, 25Sn_1800, 27Sn_1400, and 27Sn_1800, comparing both tetragonal (solid symbols) and cubic samples (open symbols). They have identical $\mu_0H_{c2}(T)$ behavior with $\mu_0H_{c2}(0.3K) = 29.1 \pm 0.2$ T. This result strongly contrasts with the widely reported depression of $\mu_0H_{c2}(T)$ for tetragonal $Nb_3Sn$ [13] [14] [15], as is shown in the comparisons of Fig. 4(b) and (c). These new samples of known, high homogeneity show for the first time that the cubic-tetragonal transformation does not control $H_{c2}$. Also of note is that the $\mu_0H_{c2}(T)$ of our bulk $Nb_3Sn$ is significantly higher than most previous reports [13][14] [15], with the exception of Foner's [14] cubic single crystals. In fact the literature finding that tetragonal $Nb_3Sn$ shows a significant depression of $\mu_0H_{c2}(T)$ comes from a sister crystal without chemical characterization [15].

There remains the issue of the normal state resistivity dependence of $\mu_0H_{c2}$. Orlando *et al* [15] noticed a crossover in on going from a "clean" film with $\rho = 8.8$ $\mu\Omega$–cm (RRR~9.5) to a rather "dirty" sample with $\rho = 36$ $\mu\Omega$–cm), $\mu_0H_{c2}(0)$ rising from ~26.5 to 29 T. Table I similarly divides our samples into a "cleaner" group with resistivities around $\rho$~10 $\mu\Omega$–cm and "dirtier" samples with $\rho$>20 $\mu\Omega$–cm. Fig. 5(a) shows that the "cleaner" samples have a lower $\mu_0H_{c2}(0.3\ K)$ ~27 T. Fig. 5(b) shows that $\mu_0H_{c2}$ and A15 phase Sn content % are much less well correlated, since samples with Sn contents of 23.3 and 24.6at.%Sn both have $\mu_0H_{c2}(0.3K) > 29$T. This result is in agreement with the expectations of standard GLAG expressions for $H_{c2}$ as shown in Orlando *et al.* [16].

The conventional view that $\mu_0H_{c2}$ is suppressed in the tetragonal state and speculations that this may explain some of the strong reversible strain sensitivity of the superconducting properties [17,18] has been derived from compilations of limited $\mu_0H_{c2}$ data taken on a wide variety of samples fabricated in different ways and of various material forms (thin films, single crystals and polycrystals). By contrast here we have varied the Sn% while maintaining the same sample quality, form and fabrication method. Multiple measurement types have directly addressed our sample homogeneity issue, leading us to conclude that all except one sample (27Sn_1400 – see Fig. 1) are indeed very homogeneous. We also found that our highest $\mu_0H_{c2}$ (0.3 K) values of 29.1± 0.2 T agree well with the highest previous measured $\mu_0H_{c2}$ values [16, 18]. We conclude that the previously reported low values of $\mu_0H_{c2}$ (0) ~21.4 T for tetragonal Nb$_3$Sn [13] [14] [15] are not determined by whether they undergo the cubic to tetragonal transformation or not. Finally we note that Cooley *et al.* [20] have shown that ball milling followed by low temperature reactions (~650°C) can strongly increase the $\mu_0H_{c2}(T)$ slope at $T_c$. Based on standard Werthamer, Helfand and Hohenberg (WHH) fitting to $\mu_0H_{c2}$ of data up to 9 T, they deduced that $\mu_0H_{c2}(0)$ could reach 35 T. However, to make our ball-milled samples homogeneous, we had to take our samples to temperatures well above 1200˚C where this valuable milling disorder anneals away. But in search of the maximum critical field, we may note the surprising result from the present binary samples is that $\mu_0H_{c2}(0)$ = 29 T is as high as any optimized Ta- or Ti-doped wire, suggesting that wires do not yet have the optimum $\mu_0H_{c2}$ possible in the system.

This work was supported by the U.S. Department of Energy under Awards No. DE-FG02-06ER54881 and DE-FG02-07ER41451. High field magnet measurements were supported by NSF/DMR-0084173 and by the State of Florida. Younjung Jo is supported by the National Research Foundation of Korea under Contracts NRF No.2010-0006377 and 2010-0020057. A. Polyanskii, W. Starch, V. Griffin, E. Choi, M. Brown and J. Collins provided valuable assistance.

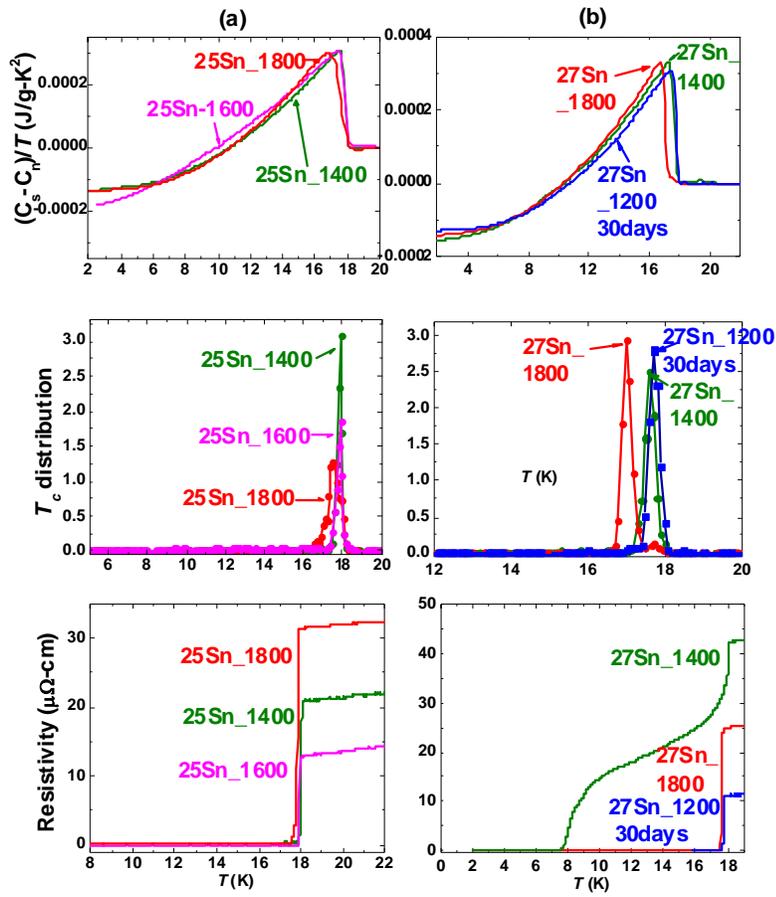

**Fig. 1** The $T_c$ plots from the resistivity and specific heat measurements for overall 25 at% (a) and 27 at% Sn samples (b). The $T_c$ distribution is obtained by deconvolution of the specific heat data in the region of the superconducting transition.

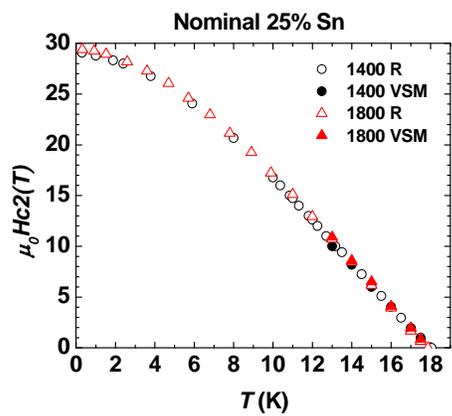

**Fig. 2** High sample homogeneity is evidenced by overlap of the percolative $H_{c2}$ deduced from the 90% point on the resistivity curves and the volumetric average $H_{c2}$ deduced from reversible magnetization measurements up to 14 T in a VSM.

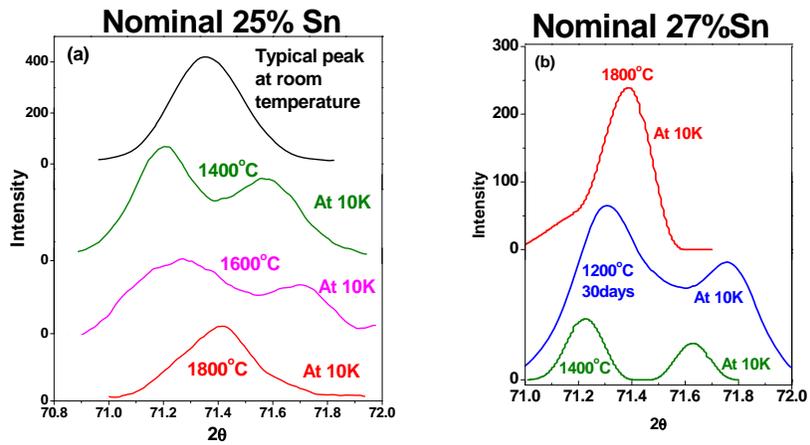

**Fig. 3** Samples with and without the signature of the low temperature tetragonal transformation shown by their low temperature XRD traces for the nominal 25%Sn (a) and 27%Sn (b) samples in all annealing conditions. Note that the transformation does not occur for samples annealed at 1800°C when the A15 Sn content falls below 24at.%.

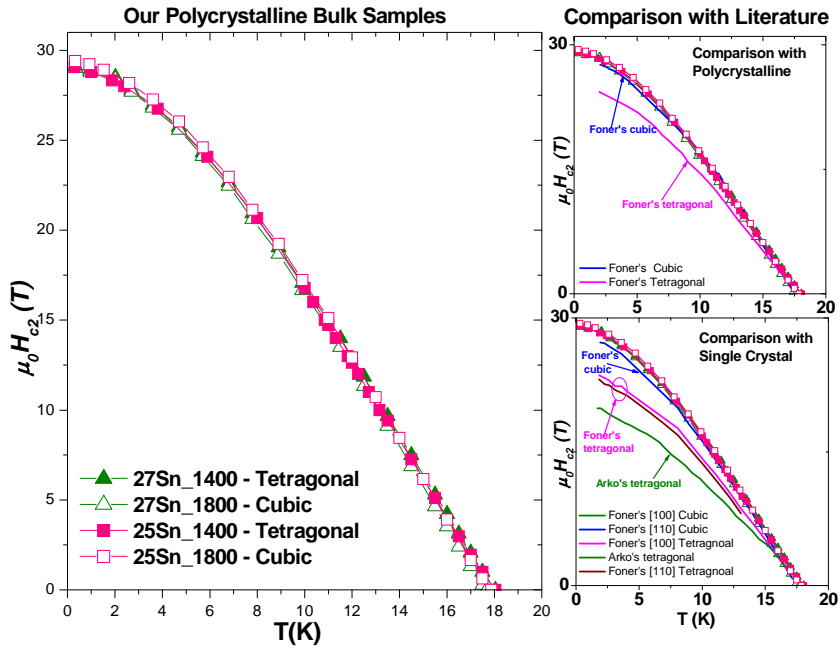

**Fig. 4** (a) Representative $\mu_0H_{c2}$ plot of our bulk Nb$_3$Sn samples. (b) Comparison to Foner's polycrystalline Nb$_3$Sn sample. (c) Comparison to Foner's and Arko's single crystal Nb$_3$Sn samples. [12] [13] [14].

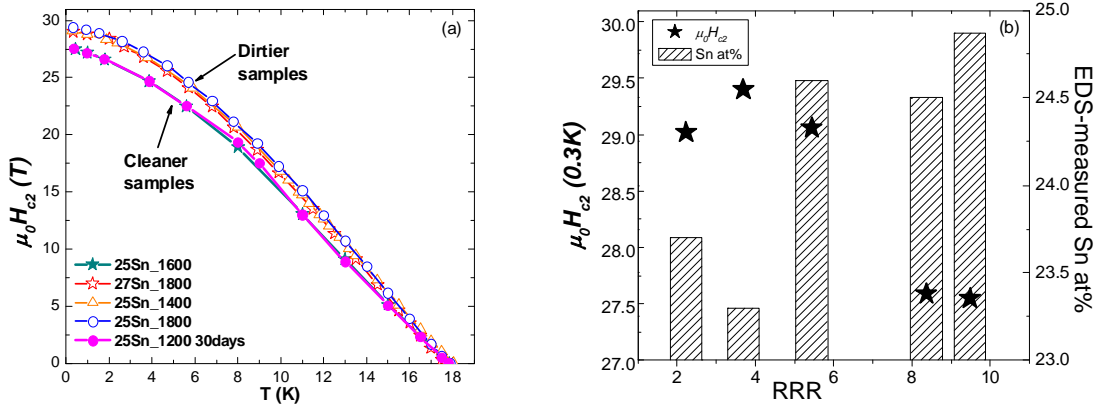

**Fig. 5** Dividing our samples into "cleaner" and "dirtier" samples following the work or Orlando *et al.* [15] we show that the "cleaner" samples lower $\mu_0H_{c2}(T)$. (a) shows the depressed $\mu_0H_{c2}(T)$ of the cleaner samples and (b) demonstrates that the $\mu_0H_{c2}(T)$ relates to the cleaner samples (lower normal resistivity and higher RRR values) rather than the EDS-measured Sn content for the Sn-rich (23.3~24.7% Sn) samples.

**Table 1** Summary of physical properties of the samples

| Sample Name | $T_c$ (K) | $a_0$ by XRD (Å) | EDS-measured Sn (at%) | RRR | $\rho$ (μΩ-cm) | $\mu_0 H_{c2}$(0.3 K) (T) |
|---|---|---|---|---|---|---|
| 25Sn_1400 | 18.05 | 5.2882±2E-4 | 24.6±0.3 | 5.44 | 20.9 | 29.1 |
| 25Sn_1600 | 17.90 | 5.2872±6E-4 | 24.5±0.4 | 8.37 | 12.7 | 27.6 |
| 25Sn_1800 | 17.87 | 5.2871±3E-4 | 23.3±0.7 | 3.69 | 31.3 | 29.4 |
| 27Sn_1400 | 18.05 | 5.2871±3E-4 | 24.6±0.2 | 1.97 | 42.5 | 29.1 |
| 27Sn_1800 | 17.67 | 5.2873±2E-4 | 23.7±0.4 | 2.23 | 24.9 | 29.0 |
| 27Sn_1200 30days | 17.81 | 5.2871±1E-4 | 24.9±0.2 | 9.48 | 9.48 | 27.6 |